\documentclass{aa}
\newcommand{\sr}{R_{\odot}}

\usepackage{color}
\usepackage{graphicx}
\usepackage{txfonts}
\usepackage{natbib}
\bibpunct{(}{)}{;}{a}{}{,}

\begin{document}
   \title{Turbulent magnetic pumping in a Babcock-Leighton solar dynamo
  model}

   \subtitle{}

   \author{G. Guerrero
          \inst{1}
          \and
          E. M. de Gouveia Dal Pino\inst{1}\fnmsep\thanks{}
          }


   \institute{Astronomy Department, Instituto de Astronomia,
     Geof\'{i}sica e Ci\^{e}ncias Atmosf\^{e}ricas\\
     Universidade de S\~{a}o Paulo,
     Rua do Mat\~{a}o 1226, S\~{a}o Paulo, Brazil\\
     \email{guerrero,dalpino@astro.iag.usp.br}
   }

   \date{}


   \abstract
  {The turbulent pumping effect corresponds to the transport of
       magnetic flux due to the presence of density and turbulence
       gradients in convectively unstable layers. In the induction
       equation it appears as an advective term and for this reason it
       is expected to be important in the solar and stellar dynamo
       processes.}
       {In this work, we have explored the effects of the turbulent
       pumping in a flux-dominated Babcock-Leighton solar dynamo
       model with a solar-like rotation law.}
       {In a first step, only vertical pumping has
  been considered through 
        the inclusion of a radial diamagnetic term in the induction
       equation. In a second step,  a  latitudinal pumping term
       has been also included
       and then, in a third step a near-surface shear has been  switched on in
       the model.}
       {The
       results reveal the importance of the pumping mechanism for
       solving current limitations in mean field dynamo modeling such
       as the storage of the magnetic flux and the latitudinal
       distribution of the sunspots. In the case that a meridional
       flow is assumed to be present only in the upper part of the
       convective zone, it is the full turbulent pumping that
       regulates both the 
       period of the solar cycle and the latitudinal distribution of
       the sunspots activity. In models that consider shear near
       the surface, a second shell of toroidal field is generated
       above $r$$=$$0.95\sr$ at all latitudes. If the full pumping is
       also switched on, the polar toroidal fields are efficiently advected
       inwards, and the toroidal magnetic activity survives only at
       the observed latitudes near the equator.
       With regard to the parity of the magnetic field, only
       models that  combine turbulent pumping with near-surface shear
       always converge  to the dipolar parity.}
       {This result suggests that, under the Babcock-Leighton approach,
       the equartorward motion of the observed magnetic activity is
       governed by the latitudinal pumping of the toroidal magnetic
       field rather than by a large scale coherent meridional
       flow. Our results support the idea that the
	 parity problem is 
       related to the quadrupolar imprint of the meridional flow on
       the poloidal component of the magnetic field and the turbulent
       pumping positively contributes to wash out this imprint.}


   \maketitle


\section{Introduction}
Flux-dominated Babcock-Leighton (FDBL) solar dynamo are mean field
models where the poloidal field is generated at the surface by the
transport and decay of bipolar magnetic regions (BMRs) which are
formed by twisted buoyant magnetic flux ropes. For this process to
occur, differential rotation must be able to develop intense
toroidal magnetic fields either at the tachocline or at the
convection zone. Numerical simulations have shown that magnetic flux
tubes with intensity around $10^4$-$10^5$ G are able to become
buoyantly unstable and to emerge to the surface to form a bipolar
magnetic region with the appropriate tilt, in agreement with the
Joy's law. One important limitation of the  scenario above is that
$10^5$ G results an energy density that is an order of magnitude
larger than the equipartition value, so that a stable layer is
required to store and amplify  the magnetic fields. This brings
another problem with regard to the way by which the magnetic flux is
dragged down to deeper layers.

\noindent In the lack of accurate observations of the flow at the
deeper layers, numerical simulations have shown that the penetration
of the plasma is restricted to only a few kilometers below the
overshoot layer \citep{gil04,rudetal05},  nevertheless, the magnetic
fields can be transported, not only downwards, but also
longitudinally and latitudinally when strong density and turbulence
gradients are present in the medium due to the turbulent pumping
\citep{zie03,dorch01}.

\begin{figure}[ht]
\centering
  \includegraphics[scale=.44]{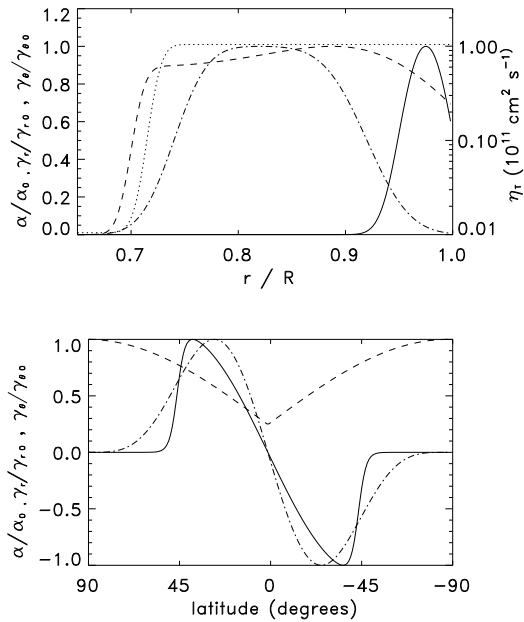}
\caption{Radial and latitudinal profiles for $\alpha$ (continuous
  line), $\eta_T$ (dotted line) and for the pumping terms $\gamma_r$ and
  $\gamma_{\theta}$ (dashed and dot-dashed lines,
  respectively). All the profiles are normalized to their maximum
  value.}\label{fig1}
\end{figure}

\noindent In axisymmetric mean field models of the solar cycle, the
effects of the turbulent pumping have been rarely considered. A
first approach showing the importance of the pumping in the solar
cycle was made by \cite{branetal92}, since then few works have
incorporated the diamagnetic pumping component in the dynamo
equation as an extra diffusive term which provides a downward
velocity \citep{kuketal01,bon02,bon06}. More recently,
\cite{kapia06} have implemented simulations of mean field dynamo in
the distributed regime, including all the dynamo coefficients,
previously evaluated in magneto-convection simulations
\citep{ossetal02,kapia06a}. They have produced butterfly diagrams
that approximately resemble  the observations. However, to our
knowledge no special efforts have been done to study the pumping
effects in the meridional plane (i.e., inside the convection zone)
or in a FDBL description. The latter has been found to be
particularly successful at reproducing most of the large scale
features of the solar cycle \citep[][hereafter GDPa,
b]{dik99,dik04,gue07a,gue07b}.\\

\noindent In this work, we explore the effects of the turbulent
pumping on a FDBL model. In a first approximation, we include the  
radial turbulent diamagnetism velocity term in the induction
equation as described by \cite{kit92}, and then in a second approach
we add the pumping terms calculated in local magneto-convection
simulations \citep{ossetal02,kapia06a}. This latter approximation
includes not only the radial but also the latitudinal contribution of
the  pumping. Finally we will also discuss the
  implications of the pumping when the near-surface radial shear layer 
  reported by \cite{cor02} is considered. In the next, we briefly
present the model, our results and then, we outline the main
conclusions of this work.  

\begin{figure}[ht]
\centering
  \includegraphics[scale = 1]{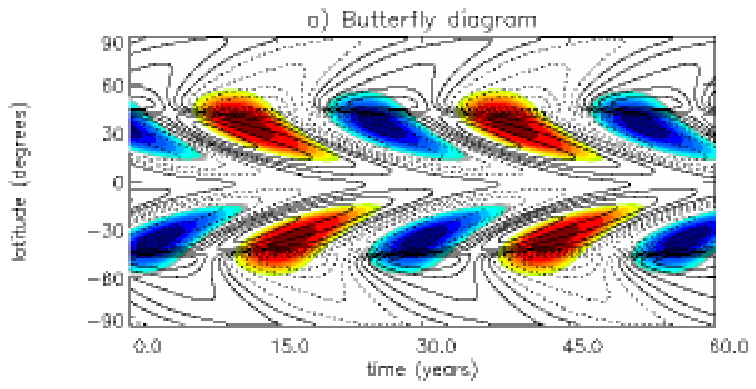}\\
  \includegraphics[width = 8cm, height=6cm]{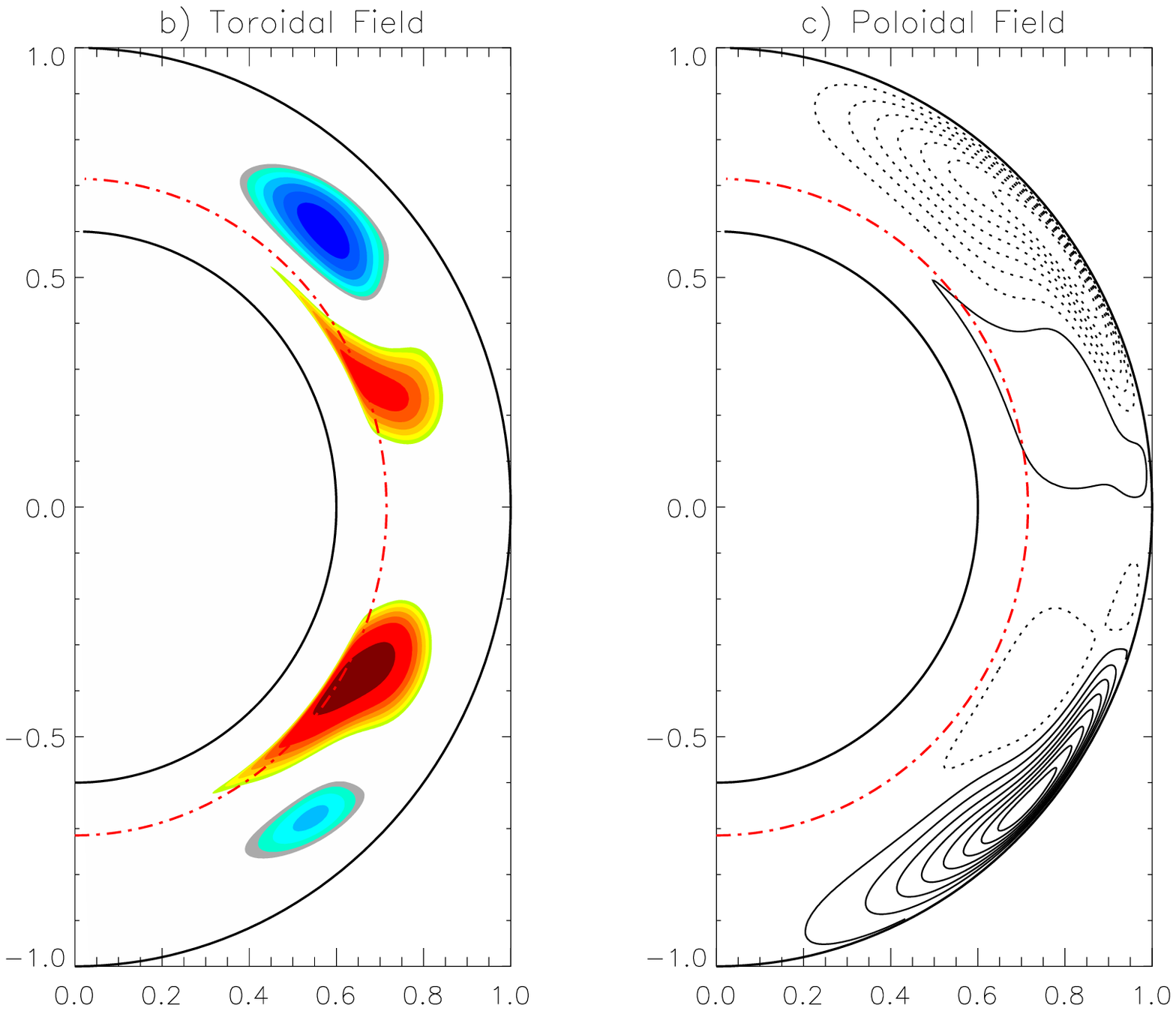}
\caption{(a) Butterfly diagram and latitudinal
  snapshots for the 
  toroidal (b) and the poloidal (c) fields. The dark (blue) and light
  (red) gray (color) scales represent positive and negative toroidal
  fields, respectively; while the continuous and dashed lines represent
  the positive and negative poloidal fields, respectively. For this
  model $T$$=$$13.6$ yr, $B_{\phi_{max}}$$=$$4.06 \times 10^4$ G and 
  $B_{r_{max}}$$=$$146.8$ G. Only toroidal fields larger than
  $2\times10^4$ (the most external contours) are shown in panels (a)
  and (b). This model has started with anti-symmetric
  initial condition (see \S5 for details.)}\label{fig2} 
\end{figure}

\section{The model}
Our model solves the mean field induction equation:
\begin{equation}\label{eq1}
\frac{\partial {\bf B}}{\partial t} = \nabla \times [{\bf U} \times
    {\bf B} + {\bf \cal{E}} - \eta_T \nabla \times {\bf B} ],
\end{equation}
where ${\bf U}$$=$${\bf u_p} + {\bf \Omega} r \sin \theta$ is the
observed velocity field, $\Omega$ is the angular velocity, ${\bf
B}$$=$$\nabla \times (A\hat{e}_{\phi}) + B_{\phi} \hat{e}_{\phi}$
are the poloidal and toroidal components of the magnetic field,
respectively, $\eta_T$ is the magnetic diffusivity and
\begin{equation}\label{eq2}
{\bf \cal{E}}={\bf \alpha} {\bf B} + {\bf \gamma} \times {\bf B},
\end{equation}
corresponds to the first order terms of the expansion of the
electromotive force, $\overline{{\bf u} \times {\bf b}}$, and
represents the action of the small-scale fluctuations over the large
scales. The coefficients of (\ref{eq2})  are the so called dynamo
coefficients. Normally, the mean field models do not consider the
second term on the right hand side of the equation, i.e, the
turbulent pumping. The first term corresponds to the alpha effect
that has been considered in several ways in the literature. A pure
Babcock-Leighton model is an $\alpha \Omega$ dynamo which represents a
large scale version of the $\alpha_{\phi \phi}$ component of  the
${\bf \alpha}$ tensor. It should resemble the buoyant rising and
twisting of strong magnetic flux tubes, reproducing events of fast
emergence of them (see the continuous lines in the diagrams of
Figure 1). We note that another source of poloidal field may exist
inside the convection zone and at the tachocline, but for our purpose
we consider only the Babcock-Leighton $\alpha$ effect since it is in 
fact, observed at the surface.

\begin{figure}[ht]
\centering
  \includegraphics[scale = 1]{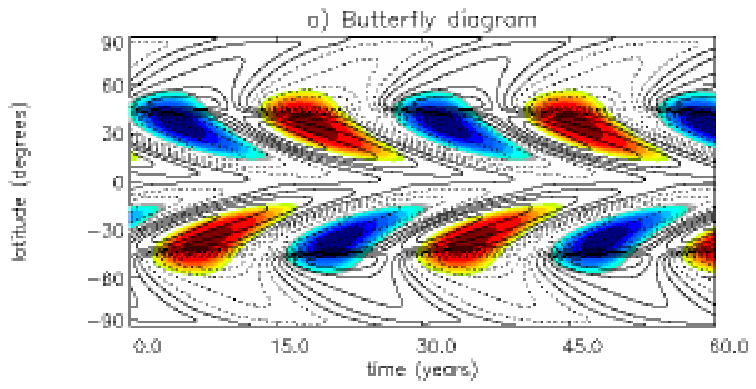}\\
  \includegraphics[width = 8cm, height=6cm]{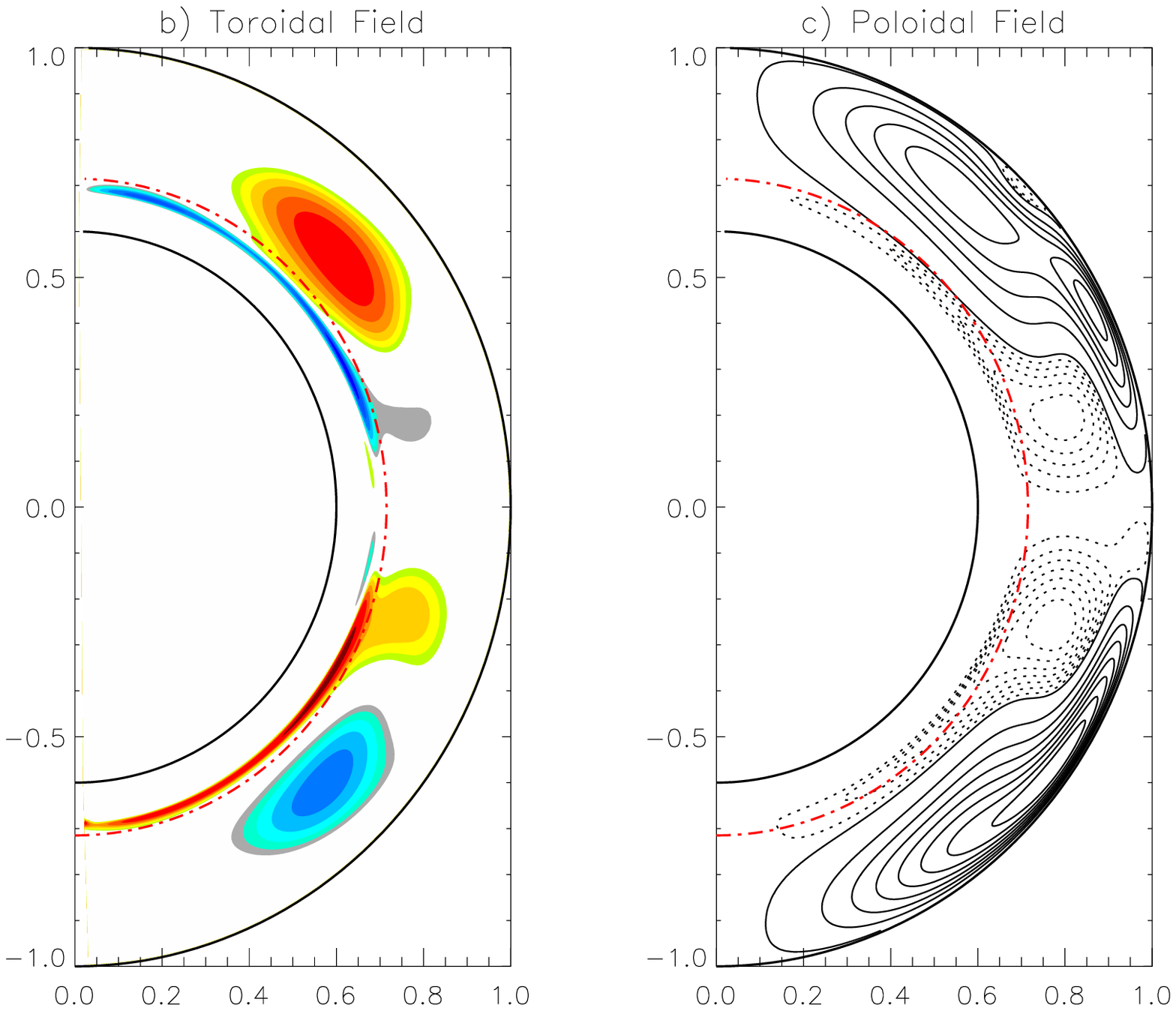}
\caption{The same as in Figs. 2 but considering here the
  radial diamagnetic term. For this model $T$$=$$13.6$ year,
  $B_{\phi_{max}}$$=$$3.9 \times 10^4$ G and $B_{r_{max}}$$=$$147.6$
  G. This model has started with anti-symmetric
  initial condition.}\label{fig3}     
\end{figure}

\noindent We solve  equation (\ref{eq1}) for $A$ and $B_{\phi}$ with
$r$ and $\theta$ coordinates in the spatial ranges $0.6\sr$$-$$\sr$
and $0$$-$$\pi$, respectively, in a $200 \times 200$ grid
resolution \cite[see details of the numerical model in]{gue04}.  

\section{Results}
The velocity field (${\bf U}$) considered in the calculations below
corresponds to the analytical profiles of eqs. (4) and (5) of
\cite{dik99}. According to the Babcock-Leighton mechanism, the alpha
term ($\alpha B_{\phi}$) is concentrated between $0.95 \sr$ and
$\sr$ and at the latitudes where the sunspots appear (see the
continuous lines in Fig. 1). Since it must produce the emergence of
magnetic flux tubes, we consider this term as being proportional to
the toroidal field $B_{\phi}(r_c,\theta)$ at the overshoot interface
$r_c$$=$$0.715\sr$. For the magnetic diffusion, we consider only one
gradient of diffusivity located at $r_c$, which separates the
radiative stable region (with $\eta_{rz}$$=$$10^9$ cm s$^{-2}$) from
the convective turbulent one (with $\eta_{cz}$$=$$10^{11}$ cm
s$^{-2}$) (see the dotted line in the upper panel of Fig. 1). The
non-dimensional parameters as defined in \cite{dik99} and employed
in the models have the following values: $R_m$$=$$U_0 \sr /
\eta_{cz}$$=$$695.5$, $C_{\Omega}$$=$$\Omega_{eq} \sr^2 /
\eta_{cz}$$=$$1.4\times10^5$ and $C_{\alpha}$$=$$\alpha_0 \sr /
\eta_{cz}$$=$$11.8$, where $R_m$ is the magnetic Reynolds number,
$U_0=$ is the maximum meridional flow velocity, $\Omega_{eq}$  is
the angular velocity at the equator, and $\alpha_0$ is the maximum
amplitude of the $\alpha$ effect.  In all the cases, the tachocline
thickness corresponds to 2\% of the solar radii and is at a radius
$R= 0.7 \sr$. 

\begin{figure}[ht]
\centering
  \includegraphics[scale = 1]{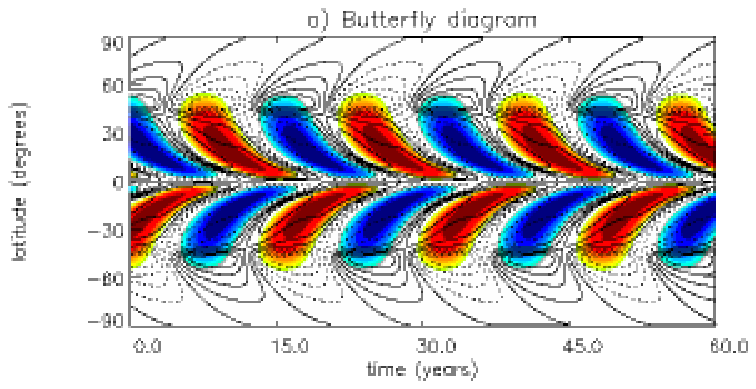}\\
  \includegraphics[width = 8cm, height=6cm]{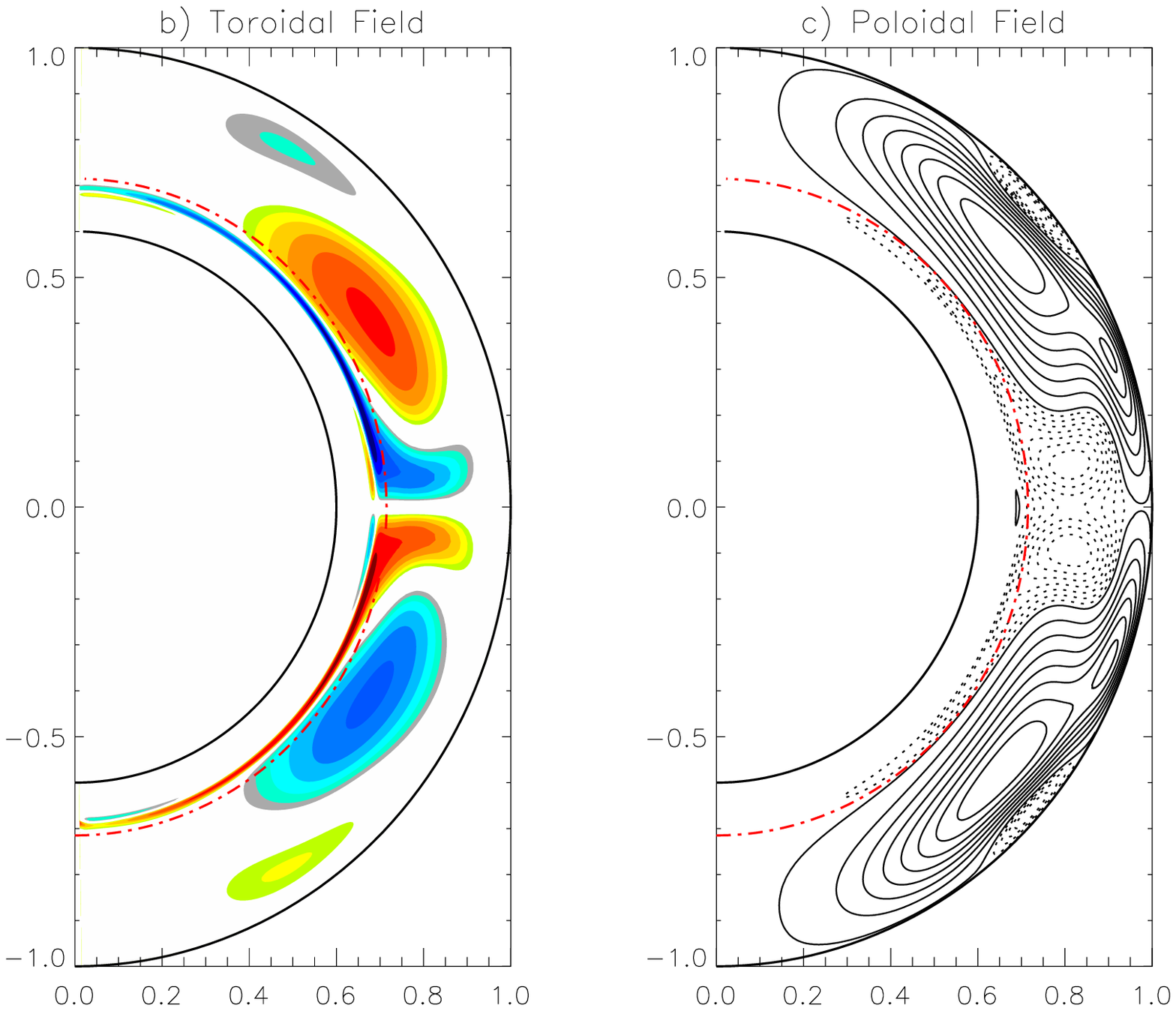}
\caption{The same as in Fig. 2 for a model with the full turbulent 
pumping terms obtained from magneto-convection simulations. For this
  model $T$$=$$8.2$ yr, $B_{\phi_{max}}$$=$$4.8 \times 10^4$ G and
  $B_{r_{max}}$$=$$155.92$ G. This model has started
    with anti-symmetric initial condition.}\label{fig4} 
\end{figure}

\noindent In Figure 2, the turbulent pumping is not considered. As
it has been reported in  GDPa and GDPb, the most important
contribution to the toroidal field comes from the latitudinal shear
term and therefore, the field that is responsible for the observed
activity begins to be formed inside the convection zone. On one hand,
it can be seen that the penetration into the stable layer is very weak,
on the other hand, the equartorward velocity is faster, so
that the time that the toroidal field has to amplify to the values
required by the magnetic flux tube simulations is probably short.
The upper panel of Fig. 2 shows the butterfly diagram
  after a transient time of $3.3  \times 10^4$ years. At this time,
  the toroidal field 
  has almost reached a quadrupolar parity (i.e. the toroidal
  field in both hemispheres has the same sign), which is in
  contradiction with the Hale's law. In the solar dynamo modelling
  this problem is known as the parity problem and we will discuss this
  subject in more detail in section \S5. In the next sections we
consider the turbulent magnetic pumping as an alternative mechanism of
penetration. The latitude of emergence of the toroidal field depends
on the stability criterion for the buoyancy \citep[e.g. see][and their
  Figs. 1 and 2 for details]{ferriz94}.

\subsection{Diamagnetic pumping}

A way to describe the diamagnetic behaviour of a non-homogeneous
plasma based on a first order smoothing approximation (FOSA) was
outlined by \cite{kit92}. If there is an inhomogeneous diffusivity
in a fluid, it causes transport of the magnetic field with an
effective velocity. We introduce this effect in the model by
changing the term $\eta_T \nabla \times {\bf
  B}$ to $\eta_T \nabla \times {\bf B} + \nabla \eta_T \times {\bf
  B}/2$  in eq. (\ref{eq1}). This new term increases the poloidal flux
that penetrates beneath the overshoot layer (Figure 3) and, as a
consequence, more toroidal field is produced at all latitudes. The
toroidal field formed inside the  convection zone penetrates the
stable layer where it is amplified by the radial shear at the
tachocline. The equartorward velocity of the magnetic flux inside
the stable layer is smaller than the velocity right above the
overshoot interface, so that in the absence of latitudinal pumping,
the toroidal field would last longer at this layer. The radial
velocity corresponding to this effect at the overshoot region is
$U_{dia}$$=$$-\nabla \eta_T /2$$\simeq$$47$ cm s$^{-1}$ when a
variation of two orders of magnitude is considered for the
diffusivity in a thin region of $0.015\sr$. 

\subsection{Full pumping}
In a convectivelly unstable rotating plasma, the magnetic field is
not advected in the vertical direction only. The diamagnetic effect
may have components in all directions. Besides, another pumping
effect due to density gradients can develop and in some conditions
can produce an upward transport that can balance  the diamagnetic
effect \citep{zie03}. Aiming at investigating a more  general
pumping advection, we will consider in this section the integration
of eq. (1) with ${\bf \cal{E}}$  given by eq. (2) and ${\bf\gamma}$
given by  both  contributions. The radial and latitudinal components
of this total ${\bf\gamma}$ were computed numerically from
three-dimensional magneto-convection simulations by
\cite{ossetal02,kapia06a}. Similarly to \cite{kapia06}, we use
the following profiles approximately fitted from the numerical
simulations:
\begin{eqnarray}
\gamma_\theta &=& \gamma_{0\theta} \left[
  1+\mathrm{erf}\left(\frac{r-0.8}{0.55}\right)\right]\\\nonumber
&\times&\left[1-\mathrm{erf}\left(\frac{r-0.98}{0.025}\right)\right]
\\\nonumber
&\times& \cos \theta \sin^4 \theta.\\
\gamma_r &=& -\gamma_{0r} \left[
  1+\mathrm{erf}\left(\frac{r-0.715}{0.015}\right) \right] \\\nonumber
&\times& \left[1- \mathrm{erf} \left(\frac{r-0.97}
  {0.1}\right) \right] \\\nonumber
&\times& \left[ \exp \left(\frac{r-0.715}{0.015}\right)^2 \cos \theta
  + 1 \right],
\end{eqnarray}
where $\gamma_{0\theta}$ and $\gamma_{0\theta}$ define the maximum
amplitudes of the pumping coefficients. The latitudinal pumping is
zero at the overshoot layer and assumes positive (negative) values at
the convection zone in the north (south) hemisphere; it is zero at the
poles and at the equator, with a 
maximum value of $100$ cm s$^{-1}$ around $\sim 15^{\circ}$ (see the
dot-dashed lines in Fig. 1). The radial pumping, $\gamma_r$, is 
negative at the convection zone, indicating a downward transport
until $r$$=$$0.7$ (i.e., below the overshoot interface) and vanishes
below this value. Its maximum amplitude is $40$ cm s$^{-1}$ (see
dashed lines in Fig. 1). We do not consider the longitudinal
contribution of the pumping because it is small compared  with the
longitudinal velocity term $u_\phi$.

\begin{figure}[ht]
\centering
  \includegraphics[height = 5 cm]{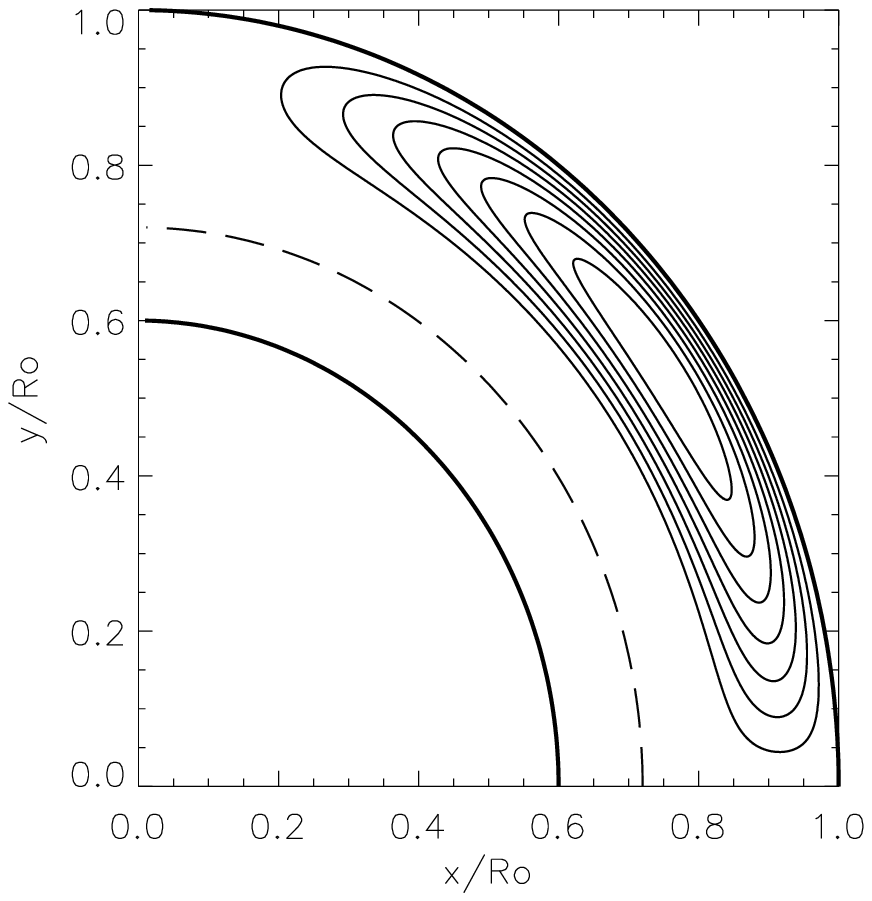}\\
  \includegraphics[scale = 1]{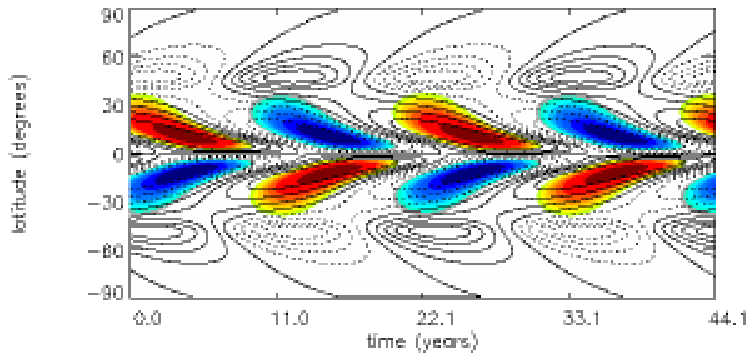}
\caption{Meridional flow streamlines and the butterfly diagram for a
model with the full pumping term, but with a shallow meridional flow 
penetration with a depth of only $0.8\sr$, $U_0$$=$$1300$ cm
s$^{-1}$, $\gamma_{\theta 0}$$=$$90$ cm s$^{-1}$ and $\gamma_{r
0}$$=$$30$ cm s$^{-1}$. For this model we obtain $T$$=$$10.8$ yr,
  $B_{\phi_{max}}$$=$$4.5 \times 10^4$ G and $B_{r_{max}}$$=$$154.9$
  G. This model has started with anti-symmetric initial
  condition.}\label{fig5}\end{figure} 

\noindent Figure 4 shows that besides advecting the magnetic fields
towards the stable regions, the pumping terms lead to a distinct
latitudinal distribution  of the toroidal fields when compared with
the previous results of Figs. 2 and 3. The turbulent and density
gradient levels present in a convectively unstable layer cause the
pumping of the magnetic field both down and equartorward, allowing
its amplification within the stable layer and its later emergence at
latitudes very near the equator. This result is important for the
dynamo modeling because it suggests that the pumping can not only
solve the problem of the storage of the toroidal fields in the
stable layer, but it can also help to provide a latitudinal
distribution that is in agreement with the observations.

\subsection{A shallow meridional flow}

As the pumping and the meridional flow are both advective
terms and in some regions inside the convection zone their radial
and  latitudinal components have the same sign, when the total
pumping is considered the period of the cycle is strongly affected.
It goes from $13.6$ yr in the models of Figs. 2 and 3, to $8.2$ yr
in the model of Fig. 4. One possibility to reproduce the solar
period is to decrease the value of the diffusivity at the convection
zone. Another possibility is to decrease the depth of penetration of
the meridional flow. This is supported by recent helioseismic
results (\cite{mitra07}) that suggest that the return point of the
meridional circulation can be at $\sim 0.95\sr$. At lower regions,
beneath $\sim 0.8\sr$, a second weaker convection cell or even a
null large scale meridional flow can exist. In Figure 5, we have
decreased the depth of penetration of the flow and found that the
period increases to the observed value at the same time that the
toroidal fields become more concentrated at lower latitudes. If we
further decrease the depth of penetration, the equatorward
concentration of the toroidal fields becomes larger and the period
longer. The reason for that is that no net magnetic flux is going
poleward at the lower regions of the convective layer since it is
going all to the equator with the pumping velocity. This result is
in agreement with \cite{ossetal02} who suggest than the equartorward
motion of the magnetic activity could not be the result of a
meridional bulk motion, but due to the latitudinal pumping of the
toroidal mean magnetic field. A parametric analysis of the
simulations performed under the conditions above results:
\begin{eqnarray}
T \simeq  181.2 \; U_0^{-0.12} \gamma_{r0}^{-0.51} \gamma_{\theta_0}^{-0.05},
\left \{ \begin{array}{l}
500 \le U_0 \le 3000 \;{\rm cm \; s^{-1}},\\
60  \le \gamma_{\theta 0} \le 140 \; {\rm cm \; s^{-1}} ,\\
20  \le \gamma_{r0} \le 120 \; {\rm cm \; s^{-1}}. \end{array} \right.
\end{eqnarray}
This indicates that the pumping terms regulate the period of the
cycle, leading to a different class of dynamo that  is
advection-dominated not by a deep meridional flow but by turbulent
pumping. Figure 5 shows a butterfly diagram with fiducial values for
 the meridional flow  which
result a period of $10.8$ yr. The agreement of this diagram  with
the main features of the solar cycle, including the phase lag
between the field components, is clear.

\section{$\Omega$ effect at the near-surface shear layer}

\begin{figure}[ht]
\centering
  \includegraphics[scale = 1]{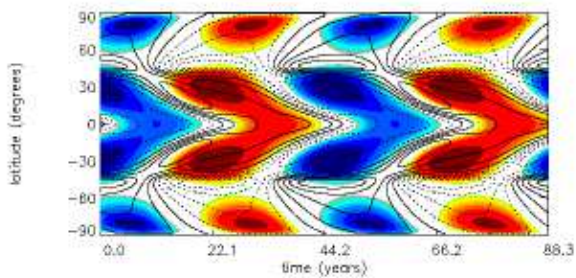}
\caption{Butterfly diagram for a model with the same
  parameters of 
  Fig. 2, but with near-surface shear layer. For this model $T$$=$$15.6$
  yr $B_{\phi_{max}}(r$$=$$0.715)$$=$$1.1 \times 10^5$,
  $B_{\phi_{max}}(r$$=$$0.98)$$=$$1.9 \times 10^4$  G and 
  $B_{r_{max}}$$=$$131.7$ G. This model has started with
  symmetric initial condition.}\label{fig6}  
\end{figure}

Helioseismology inversions have identified a second
  radial shear layer located below the solar photosphere in the upper
  $35$ Mm of the sun \citep{cor02}. It is possible, that the solar
  dynamo is operating in this region, as has been critically discussed
  by \cite{bran05}. The more attractive features 
  of an $\alpha \Omega$ dynamo operating in this region are, among
  others, the following: (i) the intensity of the magnetic flux tubes in
  this region does not need to be as large as  $10^5$ G in order to
  form sunspots  with the observed magnitudes, but $10^3$ G is
  sufficient; and (ii) with near-surface $\Omega$ effect it is
  possible to explain the 
  coincidence of the angular velocity of the sunspots in the
  photosphere with the rotation velocity at $R$$=$$0.95 \sr$ \cite[see
  Fig. 2 of][]{bran05}, as well as the apparent disconnection between 
  the sunspot and its roots \citep{koso02}. The contribution 
  of a near-surface radial shear has been investigated in
  interface-like dynamos \citep{mason02}, in distributed dynamos with a
turbulent $\alpha$ effect \citep{kapia06}, and also in advection 
  dominated dynamos \citep{diketal02}. The latter authors have 
  discarded the radial shear layer 
since it generates butterfly diagrams in which a positive toroidal
field gives rise to a negative radial field, which is exactly the
  opposite to the observed. In this section, we include the radial
  shear term in our FDBL model in order to explore the
  contribution of the pumping to this new configuration. We use the
  analytical expression given in eqs. (1)-(3) of \cite{diketal02}. The
  near-surface shear described by these equations relates a negative
  shear below $45^{\circ}$ with a positive shear above 
  this latitude \cite[see Fig. 1 of][]{diketal02} \footnote{This profile
  is slightly  different from that used by \cite{kapia06} since the 
  latter consider a negative radial shear at all latitudes.}.

\begin{figure}[ht]
\centering
  \includegraphics[scale = 1]{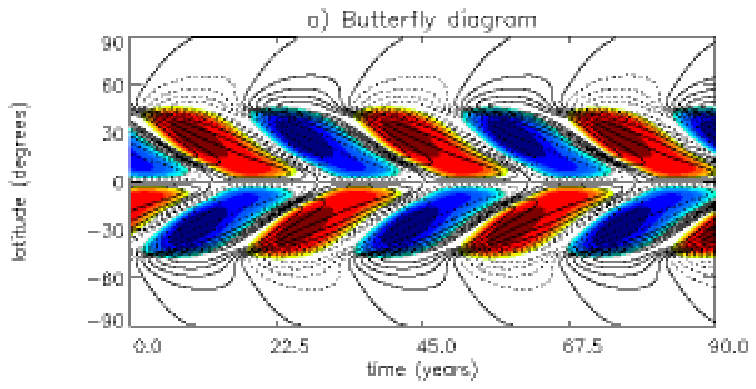}\\
  \includegraphics[width = 8cm, height=6cm]{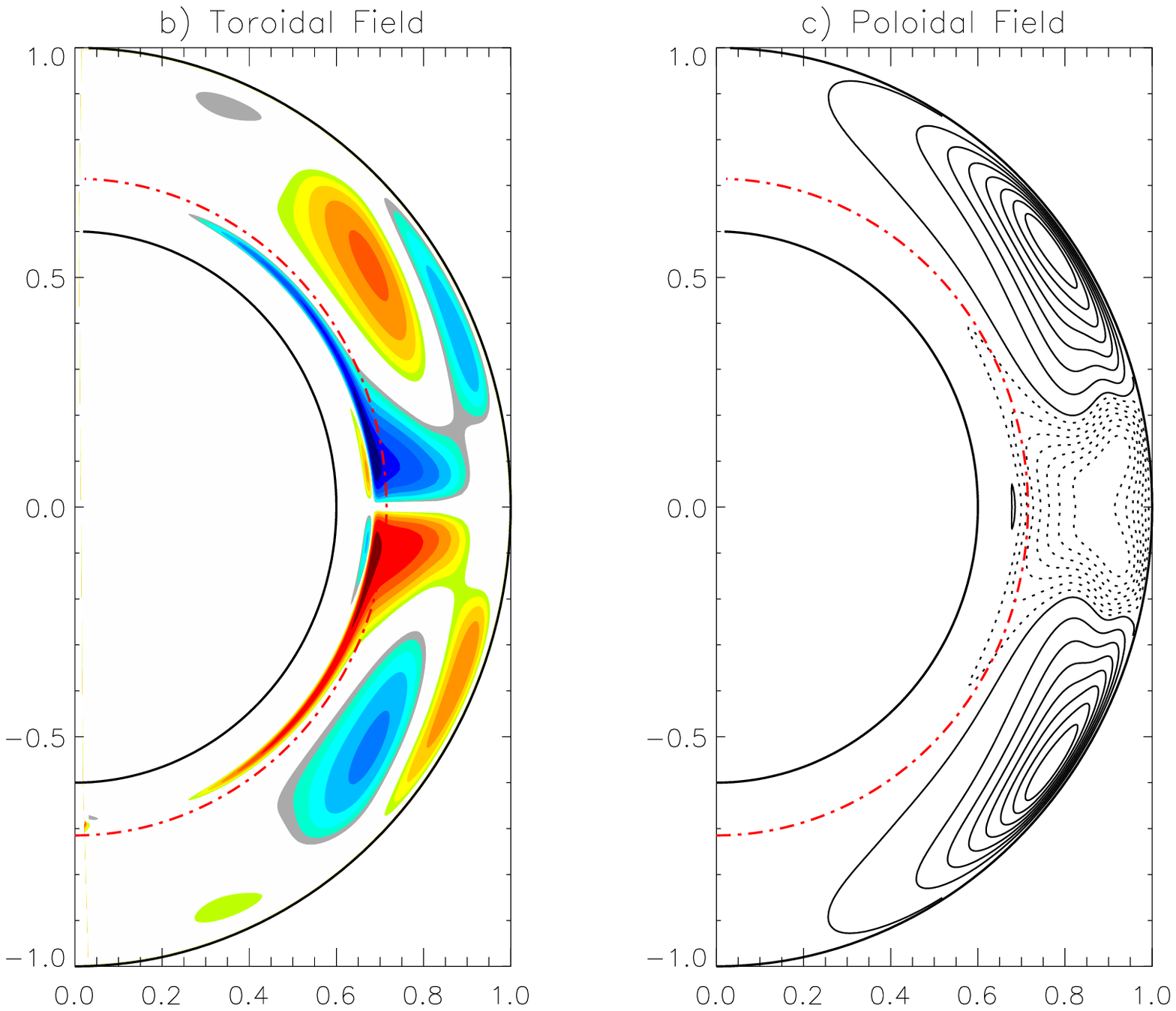}
\caption{The same as in Fig. 5 for a model with
  near-surface shear 
  action. For this model $T$$=$$16.3$ yr
  $B_{\phi_{max}}(r$$=$$0.715)$$=$$9.7 \times 10^4$,
  $B_{\phi_{max}}(r$$=$$0.98)$$=$$1.9 \times 10^4$  G and
  $B_{r_{max}}$$=$$164.4$ G. This model has started with
  symmetric initial condition.}\label{fig7}   
\end{figure}

\noindent 
With the assumption that the sunspots are
  formed in
  the upper layers, the Babcock-Leighton  poloidal source term, which
  is concentrated in the same region (above $r$$=$$0.95 \sr$), does
  not have to be non-local anymore. For the same reason, the values of
  both, the radial and the toroidal fields in the butterfly diagram
  can be taken in the same radial point ($r$$=$$0.98 \sr$). Using the
  same parameters as in the model of Fig. 2, but considering a
  near-surface shear, the results of Figure 6 show two main branches
  in the butterfly diagram. One is migrating poleward (at the high
  latitudes) and one is migrating equatorward (below $45{\circ}$. This
  result is expected if the Parker-Yoshimura sign rule
  \citep{parker55,yoshi75}  
  is considered. We note that the resulting parity is quadrupolar but
  with the correct phase lag between the fields, which is opposite to
  the obtained in \cite{diketal02}. This difference probably arises
  from the fact that we are 
  using a lower meridional circulation amplitude. Anyway, the polar
  branches are strong enough to generate undesirable sunspots close to
  the poles. The period increases to $15.6$ y, which is due to the
  fact that the dominant dynamo action at the surface goes in the
  opposite direction to the meridional flow. \\

\noindent 
In Figure 7 the same parameters as in Fig. 5 have been  
used, but this time considering the radial shear near the
surface. As the radial pumping has its maximum amplitude close to the 
poles (see the dashed line in Fig. 1), the toroidal fields created
there are efficiently pushed down 
before reaching a significant amplitude, so that only the
equatorial branches below $45^{\circ}$ survive. This 
scenario requires that the pumping be dominant over the buoyancy at
such latitudes. Also, the phase relation of $B_r
B_{\phi}$, obtained in the model of Fig. 7 seems to be the one
observed, at least at the latitude of 
activity, however, there is some overlapping between one cycle and the 
next. Results which are in better agreement with the observations 
may be achieved if the parameters are finely tuned.\\
 
\noindent 
We note that the introduction of the radial
shear 
close to the surface when a meridional flow cell penetrating down to
$r$$=$$0.71 \sr$ is considered, as in the model of Fig. 4, requires
an increase of the amplitude of $\alpha$. This result is in
agreement with that found by \cite{kapia06}
  
\section{Brief remarks on the parity problem}

Despite that it has been already explored by several
  authors, the anti-symmetry (dipolar parity) or symmetry 
  (quadrupolar parity) of the toroidal magnetic fields across the
  solar equator still constitutes one of the most challenging
  questions in the solar dynamo theory. This
  is mainly because the resulting parity in a model is very sensitive
  to a huge parameter space. The solar-like (antisymmetric) solution 
could result from the effective diffusive coupling of the poloidal
field in both hemispheres \citep{ccn04}, but it may also depend on
the position and amplitude of the $\alpha$ effect
\citep{dikgil01,bon02}, or be the result of the imprint of the
quadrupolar form of the meridional flow on the poloidal magnetic
field, as argued by \cite{char07}. Small variations in the parameter
space can switch one solution from a dipolar to a quadrupolar
one. Although the main goal of this work was not to
study the parity problem itself, but the contribution of the turbulent
  magnetic  pumping, it is interesting to take advantage of the full
  sphere integration in order to see how the pumping affects the
  parity. \\

\noindent 
All the simulations presented in the previous sections 
evolved $10^7$ time steps up to $\sim$$10^4$ years. All
started with antisymetric (A) or with symmetric (S) toroidal
magnetic field, but we have also performed tests with random (R)
fields. The parity of the solution is calculated, following
\cite{ccn04}, with the equation below: 

\begin{equation}
P(t) = \frac{\int_{-T/2}^{T/2}(B_N(t) - \overline{B}_N)(B_S(t) -
  \overline{B}_S)dt} 
{\sqrt{\int_{-T/2}^{T/2}(B_N(t) - \overline{B}_N)^2 dt}
\sqrt{\int_{-T/2}^{T/2}(B_S(t) - \overline{B}_S)^2 dt}},
\end{equation}

\noindent 
where $B_N$ and $B_S$ are the values of the
toroidal magnetic field at 
$r$$=$$0.715\sr$, and $\theta$$=$$25^{\circ}$ and $-25^{\circ}$,
respectively, $\overline{B}_N$ and $\overline{B}_S$ are their
respective temporal averages over one period. The value of $P$ should
be between $+1$ (symmetric) and $-1$ (anti-symmetric) depending on the
parity of the fields. The results of our simulations with regard the
parity can be summarized as follows:

\begin{figure}[ht]
\centering
  \includegraphics[scale = 0.5]{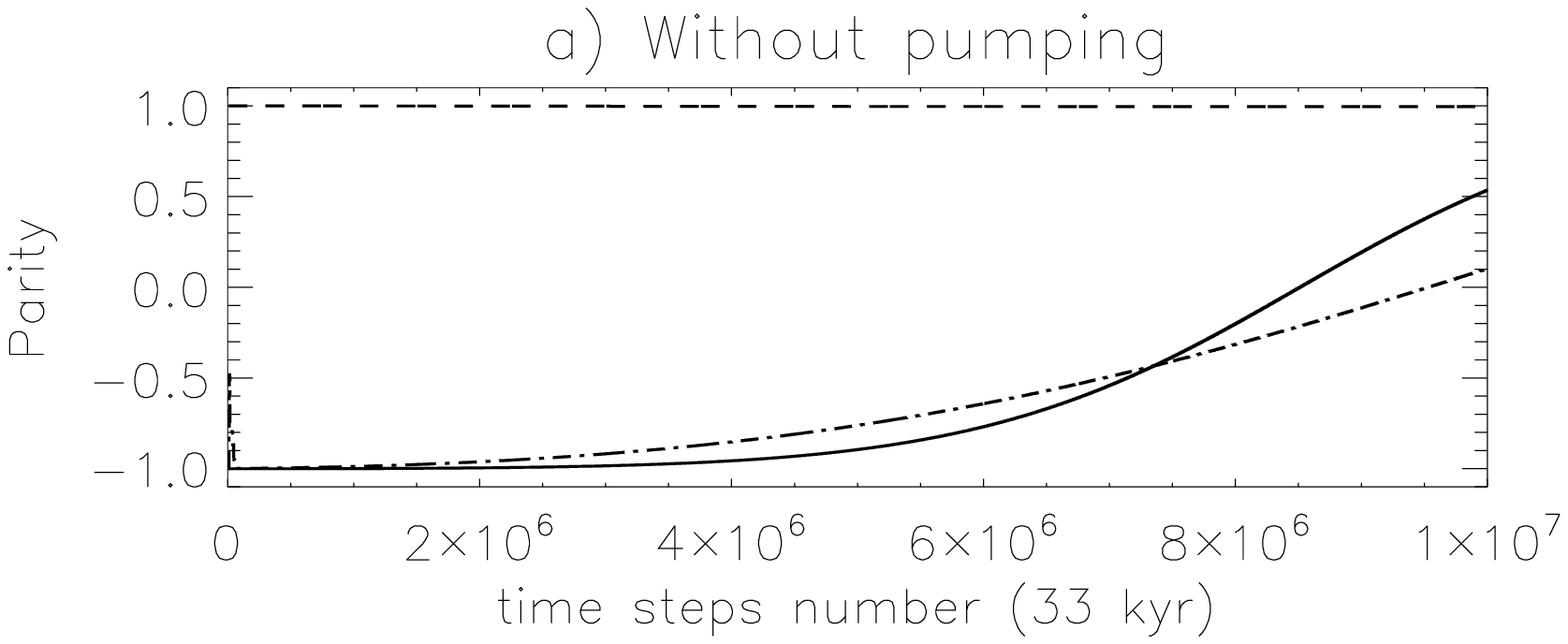}\\
  \includegraphics[scale = 0.5]{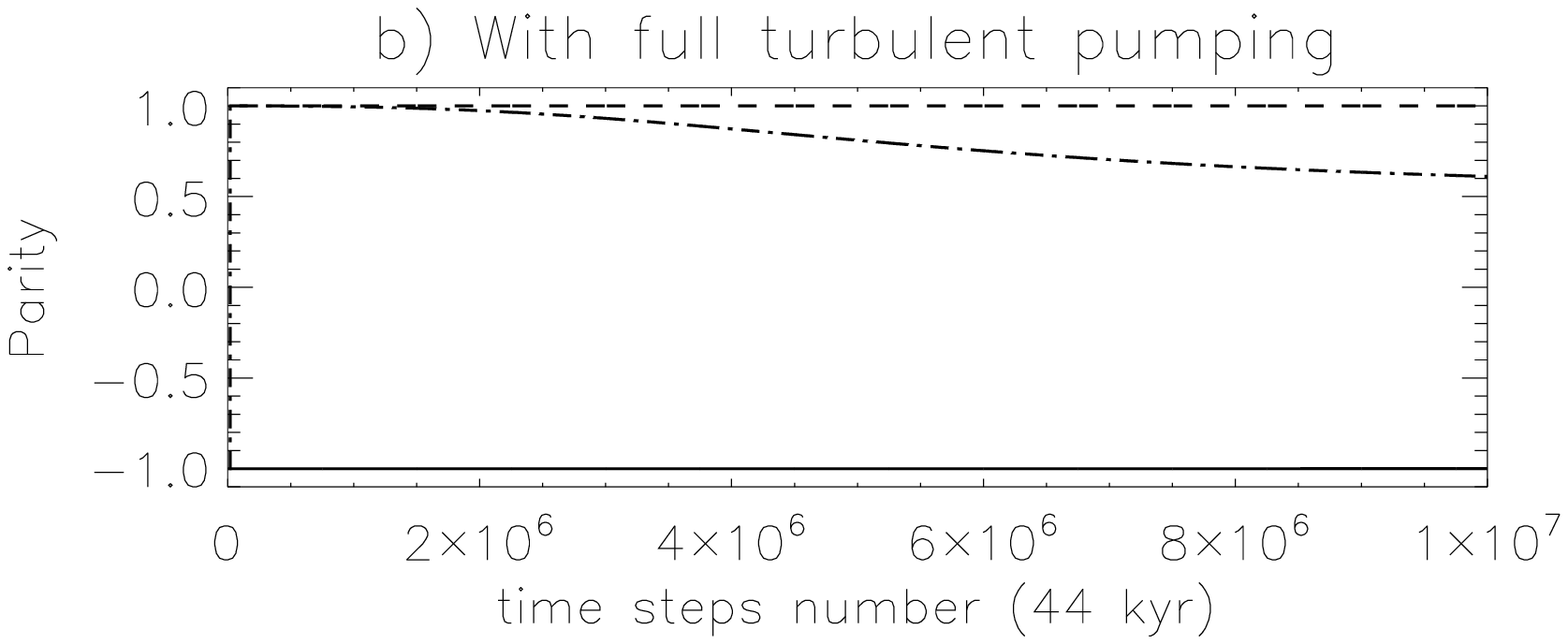}\\
  \includegraphics[scale = 0.5]{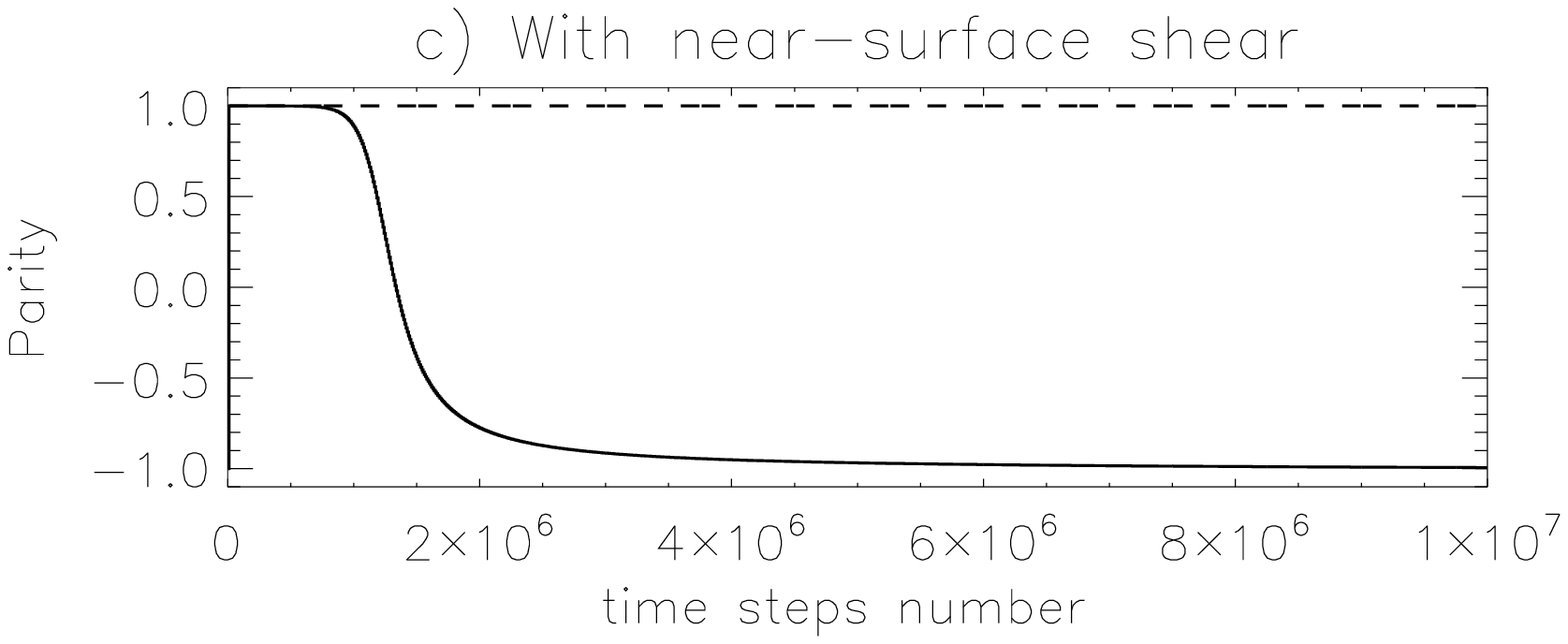}
\caption{Parity curves for the three classes of models
  considered, 
  i.e., (a) for models without pumping (as e.g., in Fig. 2); (b) for
  models with full pumping (as e.g., in Fig. 5); and (c) for models
  with near-surface shear (as e.g., in Figs. 6 and 7).  In
  the panels (a) and (b), the continuous, dashed and dot-dashed lines  
  correspond to symmetric, anti-symmetric and random initial
  conditions, respectively. In the bottom, (c), panel, the continous
  line is used for the model with turbulent pumping while the 
  dashed line is for the model without pumping.}\label{fig8} 
\end{figure}

\begin{itemize} 
\item
The models without pumping (e.g., Fig. 2), and those with diamagnetic
pumping (Fig. 3) result in quadrupolar solution. When beginning with a
dipolar initial condition they take several years before switching to
a quadrupolar solution (see Fig. 8a). This result diverges from the
one obtained by  
\cite{dikgil01} or \cite{ccn04} in which 
the change begins only after around $500$ yr. This result
indicates the strong sensitivity of the parity to the initial
parameters,  in such a way that, for example, the present parity
observed in the sun could be temporary, at least in the case that the
turbulent pumping is not relevant for the dynamo.
\item
The models with full pumping (e.g., Fig. 5) conserve
 the initial parity if this is symmetric or anti-symmetric (see Fig
 8b). When initialized with a random field, the system tends first to
 choose the quadrupolar parity, but then it tends to migrate to the  
 anti-symmetric (dipolar) parity (dot-dashed line of Fig 8b),
 suggesting that the strong quadrupolar imprint due to meridional
 circulation could be whashed out when the full turbulent pumping is
 switched on.
\item
Models with full pumping plus near-surface shear layer (e.g. Fig. 7)
tend to the dipolar parity since the first years of integration (see
continuous line of Fig. 8c). In these models we find that the
coupling of the polodial fields in both hemispheres is more
effective. This is probably due to the employment of a
local $\alpha$ term. See, for example also \cite{ccn04}, where a
near-surface $\alpha$ effect is combined with a buoyancy numerical
mechanism. They find a similar coupling. However this coupling alone is
not enough to ensure a dipolar parity (see Fig. 6 and the dashed line
of Fig. 8c). It is also necessary to eliminate the effect of the
quadrupolar shape of the meridional flow upon the poloidal magnetic
component. This can be done by the action of the pumping at the entire
convection zone (as indicated by the continuous line of Fig. 8c).   
\end{itemize}

\section{Discussion and conclusions}

We have performed 2D numerical simulations of BLFD solar dynamo
models including the turbulent pumping. 
Our first set of simulations include a solar rotation profile but
without the a near-surface radial shear layer. 
The results show that the pumping transport effect is, in fact,
relevant in solar dynamo modelling, since it can solve two important 
problems widely discussed in the 
literature: the storage of the toroidal field at the stable layer and
its latitudinal distribution. A new class of dynamo is proposed in
which the meridional flow is important only near the surface layer in
order to make the Babcock-Leighton mechanism to operate over the
toroidal fields, while in the inner layers, the advection is
dominated by the pumping velocity. Our results support the idea that
the equatorward migration of the sunspot activity is related to the
 latitudinal pumping velocity at the overshoot layer and the
convection zone. Another attractive feature of this model is that a
large coherent meridional flow is not any more required.

\noindent 
In a second set of simulations, we have
included the shear layer 
found by \cite{cor02} at the upper $35$ Mm of the sun. The results
show the formation of a second shell of strong toroidal field just
below the photosphere when the full pumping is absent. The branches of
this field obey the 
Parker-Yoshimura sign rule for a positive $\alpha$ 
effect, i.e., they move poleward at high latitudes and equatorward
below $45^{\circ}$. The role of the pumping on this kind of models is
also interesting since it reduces the amplitude of the polar toroidal
fields pushing them inwards (Fig. 7). It should be noted that these
models only work fine if a shallow meridional circulation profile is
used. When a 
deeper meridional flow going down to the tachocline is considered, a
strong $\alpha$ effect is required in order to excite the dynamo. 

\noindent 
With regard to the parity problem, our
  results show that a 
simple $\alpha \Omega$ dynamo with the $\alpha$ effect concentrated
near the surface leads to a quadrupolar parity, although the switch
from dipolar to quadrupolar 
parity takes longer than in previous studies. 
The models with full pumping conserve the initial parity, and 
when the initial condition is random, the system tends to switch to a
dipolar parity. All the models that combine full pumping with
near-surface shear prefer dipolar parity solutions too.

\noindent 
In summary, our results have demonstrated
  the importance of 
the pumping in the solar dynamo, and suggest that this effect must be
included in forthcoming studies, even in those that employ multiple
convection cells \citep{bon06,jouve07}. Besides
it decrease the influence of the meridional flow in two important
aspects: on the period of the cycle and on the latitudinal
distribution of the toroidal fields. On the other hand, our results
indicate that in the presence of full pumping there are two possible
solutions to the question on where the dynamo operates: it could be
either at the convection zone with the magnetic flux tubes
emerging from the overshoot layer, or it could be at the layers near
the surface. Both possibilities have their pros and cons, however a 
kinematic dynamo model only is not sufficient to conjecture a
definitive answer and we will explore this in more detail in
forthcoming work.

\begin{acknowledgements}
This work was supported by CNPq and FAPESP grants. G. Guerrero thanks
the MPI in Garching and the ALFA project for their kind hospitality
and support during the production of part of the present paper. We
would like to thank also to the anonymous referee for his/her
suggestions that have enriched this work.
\end{acknowledgements}

\bibliographystyle{aa}
\newpage
\bibliography{bib}

\end{document}